\documentclass[aps,prl,final,twocolumn,letterpaper]{revtex4}

\usepackage{graphicx}   
\usepackage{import}                         
\usepackage{epstopdf}
\usepackage{amsmath} 
\usepackage{float} 
\usepackage{bm}
\usepackage{amssymb}
\usepackage{quotes}
\usepackage{indentfirst}
\usepackage{color}
\usepackage{transparent}
\usepackage{dcolumn}
\usepackage{braket}
\usepackage{multirow}
\usepackage{cancel} 
\usepackage{mdframed}
\usepackage{color}
\usepackage{bm}
\usepackage{dsfont}
\usepackage{slashed}
\usepackage{soul, color} 
\soulregister\ref{7}  
\soulregister\cite{7} 
\renewcommand{\st}[1]{}

\usepackage{xr}

\makeatletter
\newcommand*{\addFileDependency}[1]{
  \typeout{(#1)}
  \@addtofilelist{#1}
  \IfFileExists{#1}{}{\typeout{No file #1.}}
}
\makeatother

\usepackage{textcomp} 
\usepackage{xifthen}
\usepackage{xcolor}
\usepackage{etoolbox}
\newboolean{togglechanges} 

\setboolean{togglechanges}{false}

\newcommand{\comment}[1]{\ifbool{togglechanges}
    {#1}  
    {\textcolor{blue}{#1}}}

\usepackage{bibentry}

\usepackage{graphicx}
\usepackage{dcolumn}
\usepackage{bm}

\usepackage{textcomp} 

\begin{document}
\rmfamily

\title{Variational processing of multimode squeezed light}

\author{Aviv~Karnieli$^{1, \S}$}
\email{karnieli@stanford.edu}
\author{Paul-Alexis Mor$^{1,\S}$}
\email{paul-alexis.mor@mines.org}
\author{Charles~Roques-Carmes$^{1,\S}$}
\email{chrc@stanford.edu}
\author{Eran Lustig$^{1}$}
\author{Jamison Sloan$^{1}$}
\author{Jelena Vu\v{c}kovi\'{c}$^{1}$}
\author{David~A.~B.~Miller$^{1}$}
\author{Shanhui~Fan$^{1}$}

\affiliation{$^{1}$ E. L. Ginzton Laboratory, Stanford University, 348 Via Pueblo, Stanford, CA 94305\looseness=-1
\\
$^{\S}$ denotes equal contribution.}



\clearpage 

\vspace*{-2em}



\begin{abstract}
Integrated multimode quantum optics is a promising platform for scalable continuous-variable quantum technologies leveraging multimode squeezing in both the spatial and spectral domains. However, on-chip measurement, routing and processing the relevant ``supermodes'' over which the squeezing resource is distributed still scales quadratically with the number of modes $N$, causing rapid increase in photonic circuit size and number of required measurements. Here, we introduce a variational scheme, relying on self-configuring photonic networks (SCN) that learns and extracts the most-squeezed supermodes sequentially, reducing both the circuit size and the experimental overhead. Using homodyne measurement as a cost function, a sparse SCN discovers the $l\ll N$ most significant supermodes using $O(lN)$ physical elements and optimization steps. We analyze and numerically simulate these architectures for both real-space and frequency-domain implementations, showing a fidelity close to unity between the learned circuit and the supermode decomposition, even in the presence of optical losses and detection noise. In the frequency domain, we show that circuit size can be further reduced by using inverse-designed surrogate networks, which emulate the layers learned thus far. Using two different frequency encoding schemes -- uniformly- and non-uniformly-spaced frequency bins -- we reduce an entire network (learning all $N$ supermodes) to $O(N)$ and even $O(1)$ modulated cavities. Our results point toward chip-scale, resource-efficient quantum processing units and demultiplexers for continuous variable processing in multimode quantum optics, with applications ranging from quantum communication, metrology, and computation.

\end{abstract}

\maketitle

\section{Introduction}

Recent years have seen a surge of interest in the field of multimode quantum optics with squeezed light \cite{fabre_modes_2020,pfister_continuous-variable_2019,armstrong_programmable_2012,roslund_wavelength-multiplexed_2014,cai_multimode_2017,gerke_full_2015,embrey_observation_2015,sharapova_bright_2018, sharapova_properties_2020,kouadou_spectrally_2023,dioum_universal_2024, arrazola_quantum_2021, madsen_quantum_2022}, with many applications for quantum technology \cite{zhuang_distributed_2018,lenzini_integrated_2018,guo_distributed_2020,roslund_wavelength-multiplexed_2014,cai_multimode_2017,ferrini_compact_2013,madsen_quantum_2022,centrone_cost_2023,armstrong_programmable_2012,arrazola_quantum_2021,pfister_continuous-variable_2019}, fundamental studies of light-matter interactions \cite{gorlach_high-harmonic_2023,heimerl_multiphoton_2024}, and nonlinear dynamics of quantum noise \cite{wright_physics_2022,rivera_ultra-broadband_2025,zia_uddin_noise-immune_2025}. One of the main drivers for the recent progress has been the promise of realizing high levels of squeezing on chip \cite{dutt_-chip_2015,zhao_near-degenerate_2020,yang_squeezed_2021, tasker_silicon_2021, zhang_squeezed_2021,arrazola_quantum_2021,nehra_few-cycle_2022,guidry_multimode_2023,lustig_quadrature-dependent_2025,shen_highly_2025}, owing to developments of low-loss, nonlinear integrated platforms for photonics. These are especially exciting for quantum optical applications employing squeezed states, ranging from quantum communication \cite{armstrong_programmable_2012,roslund_wavelength-multiplexed_2014,centrone_cost_2023}, continuous-variable quantum computation \cite{ferrini_compact_2013, lenzini_integrated_2018,pfister_continuous-variable_2019,arrazola_quantum_2021,madsen_quantum_2022, cimini_large-scale_2025} and distributed quantum sensing \cite{zhuang_distributed_2018,guo_distributed_2020}.

\begin{figure*}[t]
\centering
\vspace{-0.2cm}
  \includegraphics[width=\textwidth]{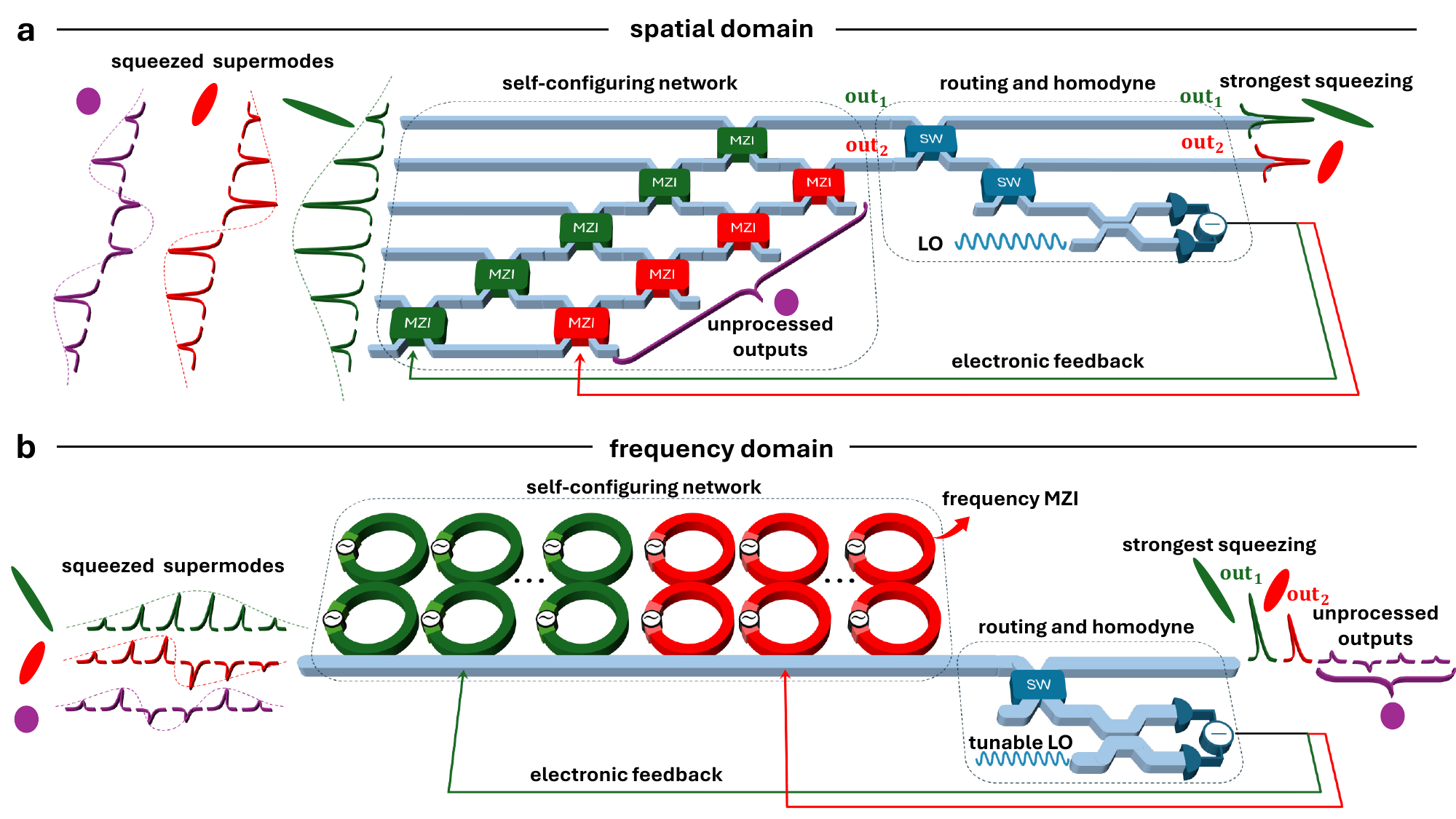}
    \caption{\small \textbf{Variational processing of multimode squeezed light using self-configuring optics.} Squeezed supermodes are processed by a self configuring network. The network comprises a mesh of Mach-Zehnder interferometers (MZIs) in a layered structure, where different layers are color-coded. Each layer has a single output port, while the rest of the ports are fed to the next layer. Output ports are routed through a layer of cross/bar switches (SW) to the homodyne measurement stage with a CW local oscillator (LO). Consecutive learning of each layer's MZI parameters is performed by optimizing over the homodyne signal (as in Fig. 2), while parameters are updated with electronic feedback. Once learning has converged, the output port of layer $i$ is guaranteed to carry the $i$-th most squeezed supermode, and the MZI parameters of layer $i$ correspond to the expansion coefficients of supermode $i$. For sparse networks (where the number of layers is smaller than the number of modes), the remaining unprocessed outputs carry the lowest squeezing, and span a subspace orthogonal to the discovered supermodes. If one wishes to find the first $l\ll N$ dominant supermodes, then the number of physical elements required is $O(lN)$. \textbf{a} Implementation in the spatial domain. Modes are encoded in the path degree of freedom (e.g., waveguide number), and MZI meshes are physically implemented on chip. \textbf{b} Implementation in the frequency domain. Modes are now encoded using frequency bins. With the equivalent of a frequency-domain MZI scattering element coupled to a waveguide \cite{hu_-chip_2021}, meshes are implemented in synthetic dimension, and the CW LO can be tuned to a specific frequency bin being measured.}
    \label{fig:concept}
    \vspace{-0.3cm}
\end{figure*}

Squeezing in multimode systems is often distributed across a linear superposition of complex degrees of freedom, such as spatial, spectral and temporal modes, forming so-called squeezed ``supermodes'' \cite{fabre_modes_2020,pfister_continuous-variable_2019}. The squeezing resource, therefore, can only be measured and utilized efficiently after the complex structure of such supermodes is known. For this reason, it is imperative to have experimental methods for decomposing an input multimode squeezed state into its supermode components. 

Current measurement techniques \cite{armstrong_programmable_2012,roslund_wavelength-multiplexed_2014,kouadou_spectrally_2023,embrey_observation_2015,volpe_multimode_2020, sharapova_bright_2018,shaked_lifting_2018, sharapova_properties_2020,amooei_efficient_2025} generally rely on direct quantum state tomography using homodyne detection (HD) \cite{fabre_modes_2020} with a local oscillator (LO) shaped into arbitrary superpositions of modes using, e.g., off-chip pulse shapers \cite{armstrong_programmable_2012,roslund_wavelength-multiplexed_2014,kouadou_spectrally_2023}, spatial light modulators and masks \cite{embrey_observation_2015,volpe_multimode_2020} or nonlinear interferometers \cite{sharapova_bright_2018,shaked_lifting_2018,nehra_few-cycle_2022}; other techniques employ direct measurement of spatial \cite{sharapova_properties_2020,amooei_efficient_2025} or spectral \cite{presutti_highly_2024} intensity correlations. On-chip implementations \cite{lenzini_integrated_2018,arrazola_quantum_2021, barral_versatile_2020,zhu_large-scale_2024,aghaee_rad_scaling_2025,dioum_universal_2024}, on the other hand, usually employ quantum processing units (QPUs) comprising programmable meshes of Mach-Zehnder interferometers (MZIs) \cite{harris_linear_2018,wang_integrated_2020,bogaerts_programmable_2020} to implement a general unitary transformation on the basis of modes, followed by a measurement stage at the output. These architectures are particularly important if we wish not only to measure but also to generate, shape, and route squeezed supermodes for further quantum communication, computation, and sensing applications. 

However, there remains a major challenge to scaling up multimode quantum information processing implemented on-chip. For quantum states encoded in $N$ modes, QPUs require $O(N^2)$ physical elements to fully determine all the supermodes. Furthermore, unless simplifying assumptions can be made on the symmetry of supermodes \cite{roslund_wavelength-multiplexed_2014, kouadou_spectrally_2023}, one generally needs to tomographically reconstruct the entire state to find even a single supermode, which requires $O(N^2)$ measurement steps in the worst case. This quadratic scaling constrains not only spatial meshes of integrated MZIs \cite{wang_multidimensional_2018, wang_integrated_2019,bogaerts_programmable_2020}, but also integrated QPUs implemented in the frequency domain \cite{lu_frequency-bin_2023,dioum_universal_2024} using spectral beam splitters \cite{hu_-chip_2021}.

Here, we propose alternative schemes for measuring, processing, and decomposing multimode squeezed states in the spatial and spectral domains, relying on variational principles. Our method employs self-configuring networks (SCNs) \cite{miller_self-configuring_2013,miller_analyzing_2020,seyedinnavadeh_determining_2024,roques-carmes_measuring_2024,roques-carmes_automated_2025, noauthor_stability_nodate, hamerly_accurate_2022, bandyopadhyay_hardware_2021}, a layered architecture for photonic circuits that allows automated sequential decomposition of an optical input into its eigenmodes. We show that by maximizing over a homodyne detection signal at the output of each layer of the SCN (see Fig. 1), with respect to the MZI variables of the layers, we can sequentially find the most squeezed supermodes in the system, while routing their respective squeezing into corresponding output ports. This permits favorable scalings with the number of modes, in both the quantity of on-chip physical elements and the number of iterations required for finding the dominant supermodes. Note, too, that this separation into supermodes by self-configuration does not rely on any calibration of the MZIs nor on their perfection. We then propose and analyze different architectures realizing these variational processors in both the spatial and the spectral domains, where the use of the synthetic frequency dimension \cite{lin_synthetic_2018} enables us to compress the spatial footprint of the \textit{entire} circuit even further down to $O(N)$ and even $O(1)$ physical elements. Our approach paves the way towards scalable on-chip processing of high-dimensional squeezed states, with applications to quantum communication, sensing, and computation.


\section*{Variational processing of multimode squeezing} 
\textit{The Bloch-Messiah decomposition.} Multimode squeezed states belong to a broader family of multimode Gaussian states, which are quantum states of light whose phase-space representation admits a multi-dimensional Gaussian form \cite{ferraro_gaussian_2005,fabre_modes_2020}. These states can be completely characterized by a mean vector and covariance matrix of the multimode quantum field quadratures. For mode $i$, represented by the photon annihilation operator $a_i$, there are two quadratures: $x_i=a_i+a_i^{\dagger}$ and $p_i=(a_i-a_i^{\dagger})/i$. Put together, these quadratures form a $2N$-long quadrature vector $\vec{q}=(x_1,x_2,...,x_N,p_1,p_2,...,p_N)$ for $N$ modes in the system. The mean vector $\braket{\vec{q}}$ corresponds to the phase-space displacement of the field, whereas the covariance matrix is given by the $2N$-by-$2N$, positive semidefinite and symmetric matrix  
\begin{equation}
    \Gamma = \frac{1}{2}\braket{\delta\vec{q}\delta\vec{q}^T + (\delta\vec{q}\delta\vec{q}^T)^T},
    \label{eq:covariance_definition}
\end{equation}
where $\delta\vec{q} =\vec{q}-\braket{\vec{q}}$.

Processing multimode squeezed states (as well as other Gaussian states) then involves the measurement and subsequent decomposition of the covariance matrix, which encodes the entire information about the squeezed supermodes. Owing to the Bloch-Messiah decomposition (BMD) \cite{fabre_modes_2020}, and in the presence of mode-independent photon loss, we are able to decompose the covariance matrix into a diagonal form via

\begin{equation}
    \Gamma = (1-p)OK^2 O^T + p I ,
    \label{eq:BMD}
\end{equation}
where $O = [\mathrm{Re} U, -\mathrm{Im} U; \mathrm{Im} U, \mathrm{Re} U]$ is a $2N$-by-$2N$ real orthogonal matrix corresponding to a general $N$-by-$N$ unitary mode mixing transformation $U$, $K = \mathrm{diag}(e^{r_1},e^{r_2},...,e^{r_N},e^{-r_1},e^{-r_2},...,e^{-r_N})$ is a diagonal squeezer, with $r_j$ denoting the $j$-th squeeze parameter, $I$ denotes the identity matrix, and $p$ is the photon loss probability. The fact that the BMD corresponds to an orthogonal diagonalization $\Gamma = O\Gamma_DO^T$ of the covariance matrix means that it could be realized using passive optical circuits that implement unitary mode mixing. Once the BMD of Eq. (\ref{eq:BMD}) is known, the complex coefficients of the $i$-th supermode ($i=1,...,N$) can be inferred from the $i$-th row of the orthogonal matrix $O$, and the amount of anti-squeezing (respectively, squeezing) carried by this supermode can be read from the $i$-th (respectively, $i+N$-th) diagonal element of $\Gamma_D$. For a more comprehensive overview of the covariance matrix, symplectic transformations, and the BMD \cite{fabre_modes_2020}, see SI Section S1A-B.

For a highly-multimode system, an important and desirable aspect of BMD using on-chip photonic circuits is not only reconstructing and measuring the most significant supermodes (i.e., the ones that carry the strongest squeezing), but also \textit{routing} their respective squeezed quadratures into different output ports for further use. We will now show, using self-configuring architectures, how this procedure can be performed sequentially and automatically, where the photonic circuit discovers the most squeezed supermodes first. We do this for both the spatial (Figure 1a) and the frequency (Figure 1b) domains, as discussed below. 

\begin{figure*}[t]
\centering
\vspace{-0.2cm}
  \includegraphics[width=\textwidth]{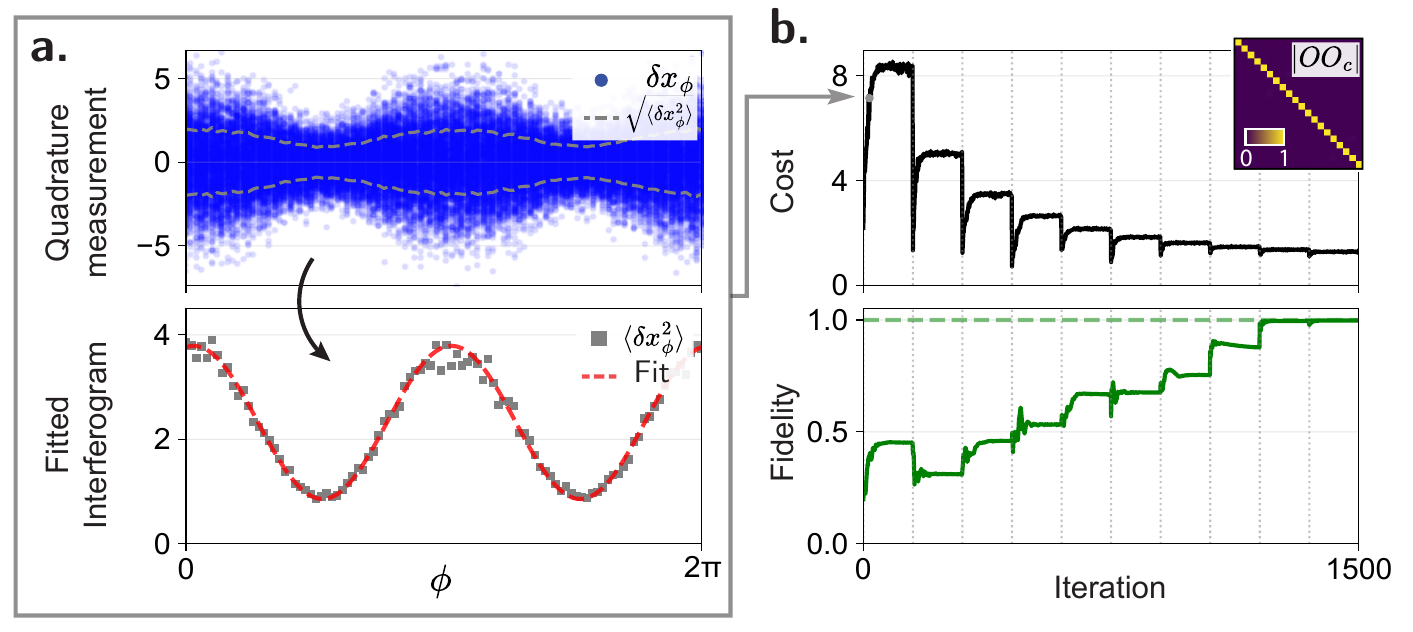}
    \caption{\small \textbf{Numerical simulation of variational learning of multimode squeezed light.} \textbf{a} Simulated quadrature measurement, and the corresponding homodyne interferogram signal for the quadrature variance. The interferogram is fitted according to Eq. (\ref{eq:homodyne}) and the cost function extracted according to SI Section S2.B. \textbf{b} Plotted cost function (Eq. (\ref{eq:Rayleigh quotient})) and overall Hilbert-Schmidt fidelity $\mathrm{Tr}|OO_C|/2N$ as a function of iteration number, where $|\cdot|$ denotes element-wise absolute value. The inset shows the product $|OO_C|$ to reflect the high fidelity between $O_C$ and $O^T$.  }
    \label{fig:Simulation}
    \vspace{-0.3cm}
\end{figure*}

\textit{Self-configuring architecture in the spatial domain.} The proposed self-configuring architecture for implementing our variational processors of multimode squeezed states in the spatial domain is depicted in Fig. 1a. The spatial bins are encoded in waveguide modes, and the photonic circuit is implemented through a mesh of MZIs. Each MZI in the spatial domain comprises two $50/50$ beam-splitters, and two electrically controllable phase shifters (e.g., using thermo-optical control), phases of which we denote by $\theta_j$ and $\varphi_j$ for the $j$-th MZI. Self-configuring networks \cite{miller_self-configuring_2013,miller_analyzing_2020,bogaerts_programmable_2020} comprise a cascade of layers of such MZI elements, with a specific topology: every input port has exactly one path through MZI blocks to the output port of the layer \cite{miller_analyzing_2020}. Examples include diagonal layers (depicted in Fig. 1a) as well as binary tree layers \cite{miller_analyzing_2020} and combinations thereof. Recently, such networks have been proposed \cite{roques-carmes_measuring_2024,roques-carmes_automated_2025} and experimentally demonstrated \cite{seyedinnavadeh_determining_2024} to sequentially find eigenstates of classical communication channels \cite{seyedinnavadeh_determining_2024}, partially coherent light \cite{roques-carmes_measuring_2024} and Schmidt modes of entangled photon pairs \cite{roques-carmes_automated_2025}. We now employ this concept to process multimode squeezed states, where we make use of a variational principle over homodyne measurements to allow the network to automatically and sequentially learn the most significant supermodes and route them to separate output ports. 

The input squeezed light comprises unknown supermodes (green, red, magenta envelopes in Fig. 1a), dispersed over a discrete set of waveguide modes. The input state enters a self-configuring network: in this example, the number of bins is $6$ whereas the network consists of two diagonal layers of MZIs. The output of each layer is the top right port in the spatial implementation (Fig. 1a). After propagating through the network, the output port of layer $1$ (denoted as $\mathrm{out_1}$ in Fig. 1a) is routed to a homodyne measurement stage using selection switches (SW), comprising MZI elements set to either totally transmit ("bar") or reflect ("cross") the input. The homodyne measurement is optimized (as will be described below) with respect to the MZI parameters of layer 1 until convergence, where electronic feedback is used to update the layer parameters.

Once the learning of the first layer has converged, the light in output port 1 is guaranteed to carry the squeezed quadratures of the most squeezed supermode, as explained below. The process can now be repeated to learn the second layer, cascaded to the first, where the parameters of layer 1 remain fixed. The output of the second layer will then carry the squeezed quadratures of the second-most squeezed supermode. If the network is sparse (the number of layers $l$ smaller than the number of modes $N$, as in the example in Fig. 1), then all spatial modes orthogonal to the first $l$ supermodes will remain unprocessed and routed to unused output ports. A key advantage made possible by the self-configuring architecture is, therefore, its ability to harvest the  strongest squeezing resource encoded in the first $l$ supermodes using sparse photonic circuits consisting of $O(lN)$ physical elements.             

\textit{Self-configuring architecture in the frequency domain.} We propose a similar architecture for self-configuring networks in the frequency domain, as depicted in Fig. 1b. The input light travels in a single spatial mode in a waveguide, and processed via scattering off electro-optically modulated micro-ring cavities side-coupled to the waveguide. The modes are encoded in frequency bins, which correspond to the resonance frequencies of the integrated ring resonators used to generate the squeezed light \cite{guidry_multimode_2023,lustig_quadrature-dependent_2025}. In the spectral implementation, the photonic circuits are built from units of integrated frequency-bin MZIs \cite{hu_-chip_2021,lu_frequency-bin_2023}, each comprising two modulated rings that can selectively couple pairs of frequency bins through a scattering process \cite{hu_-chip_2021,lu_frequency-bin_2023} (see Fig. 1b). As in the spatial domain, the $j$-th frequency-domain MZI is controllable by two degrees of freedom $\theta_j$ and $\varphi_j$ which correspond to the amplitude and phase of the electro-optic modulation of the MZI. Such integrated frequency-domain MZIs have been recently demonstrated experimentally\cite{hu_-chip_2021}. Their operation principle relies on a mode splitting between two coupled ring cavities \cite{hu_-chip_2021}, where the splitting should correspond to the frequency-bin spacing of the input (e.g., the free spectral range of the cavity generating the squeezed light).  

The self-configuring layers are then built by cascading frequency-domain MZIs, as depicted in Fig. 1b for a diagonal layer, where the $j$-th MZI couples the $j$-th and $j+1$-st frequency bins. An input multimode squeezed light enters the network and scatters off the self-configuring network, and the optimization is performed by routing the the entire output to the homodyne stage. To learn the first layer, a frequency-tunable continuous-wave (CW) LO is set to the first  (e.g., lowest) frequency bin, corresponding to the output port of the first layer (and, whenever frequency bin $j$ is optimized on, the LO frequency can be tuned to $\omega_j$). The output is then optimized according to the protocol discussed below, and electronic feedback is used to update the layer parameters.

We note that whenever considering modes encoded in the frequency domain, the modes are not strictly discrete. In fact, the quadratures can also vary as a function of a frequency detuning $\omega$ relative to the center frequency of the $i$-th frequency bin, such that $x_i(\omega) = a_i(\omega)+a_i^{\dagger}(-\omega)$ and $p_i(\omega) = (a_i(\omega)-a_i^{\dagger}(-\omega))/i$. These are in fact non-Hermitian operators that form a complex-valued covariance matrix $\Gamma(\omega)$, supporting morphing supermodes \cite{dioum_universal_2024,gouzien_morphing_2020} that vary nonuniformly with $\omega$. In the ensuing analysis, unless stated otherwise, we shall make two assumptions: first, we consider a region of $\omega$ smaller than the squeezing bandwidth around each resonance, thus effectively analyzing the decomposition of the real-valued covariance matrix $\Gamma\equiv\Gamma(0)$. Second, scattering elements that we use for optical processing in the frequency domain (such as cavities side-coupled to a waveguide) have a larger bandwidth around each resonance compared to the squeezing bandwidth of the quantum light. A unitary operator $U(\omega)$ associated with such elements acts approximately independent of $\omega$ within the relevant spectral range, and we can thus simplify $U=U(0)$. A full frequency-dependent treatment of our model is detailed in the SI section S1C.

\textit{Variational optimization using homodyne measurements.} We now detail the variational optimization procedure the network uses to learn the parameters of each self-configuring layer using homodyne measurements. Homodyne detection is a fundamental tool for characterizing quantum states of light in phase space \cite{lvovsky_continuous-variable_2009,fabre_modes_2020}, and was recently implemented on chip \cite{tasker_silicon_2021}. When a quantum light field under investigation interferes with a coherent state LO of amplitude $\alpha$ in a balanced beam splitter (BS), the intensities of the output ports of the BS are measured using photodetectors, and their difference measures the operator $N_- (\phi)=2|\alpha|(\cos \phi x+\sin \phi p)\equiv 2|\alpha|x_{\phi}$ where $\phi$ is the controllable phase of the LO. For Gaussian states  it suffices to measure the first two moments of $N_-(\phi)$: Measuring $\braket{N_-(\phi)}$ reveals $\braket{x_{\phi}}$, while measuring the variance of $N_- (\phi)$ gives us the variance in the quadrature $x_{\phi}$, or $\braket{\delta x_{\phi}^2}=\braket{x^2_{\phi}}-\braket{x_{\phi}}^2$. When the self-configuring network is learning its $i$-th layer, the homodyne measurement is performed over the $i$-th output port of the photonic circuit (ports $\mathrm{out}_i$ in Fig. 1a-b), implementing a unitary $U_c$ (correspondingly, an orthogonal transformation $O_c$), and we have that (see SI Sections S2.A-B for derivation)
\begin{equation}
\begin{split}
    \braket{\delta x_{\phi,i}^2} &= \cos^2 \phi \left[O_c\Gamma O^T_c\right]_{i,i} + \sin 2\phi \left[O_c\Gamma O^T_c\right]_{i,i+N}  \\ &+\sin^2 \phi \left[O_c\Gamma O^T_c\right]_{i+N,i+N}.
\end{split}
    \label{eq:homodyne}
\end{equation}
From the homodyne interferogram as a function of $\phi$ we can then extract a cost function for learning the $i$-th layer, in the form of the Rayleigh quotient (see SI Sections S2.A-B for derivation):
\begin{equation}
\begin{split}
\mathcal{C}[\vec{o}_c^{(i)}] = [O_c\Gamma O_c ^T]_{ii} = \frac{\vec{o}_c^{(i)T} \Gamma \vec{o}_c^{(i)}}{\vec{o}_c^{(i)T} \vec{o}_c^{(i)}}
\end{split}
    \label{eq:Rayleigh quotient}
\end{equation}
where $\vec{o}_c^{(i)}$ denotes the $i$-th column of $O_c$. Experimentally, this quantity is found from the recorded homodyne interferogram by either a Fourier transform or a least-square fit to a sinusoidal function according to Eq. \ref{eq:homodyne}, as detailed in Section S2.B.

Using the variational theorem and the cascading property of the self-configuring network described above, we show in the SI, Section S2.A that the maximum of this quantity in Eq. (\ref{eq:Rayleigh quotient}) corresponds to the $i$-th largest eigenvalue of $\Gamma$ in Eq. \ref{eq:covariance_definition}:
\begin{equation}
\begin{split}
&\mathrm{max} (\mathcal{C}[\vec{o}_c^{(i)}]) = (1-p)e^{2r_i} + p, \\
&\mathrm{argmax} (\mathcal{C}[\vec{o}_c^{(i)}]) = \vec{o}^{(i)}, \\
\end{split}
    \label{eq:maximum}
\end{equation}
with $r_1\geq r_2\geq...\geq r_i \geq ...\geq r_N$ are the ordered squeezing parameters, and $\vec{o}^{(i)}$ is the $i$-th column of $O^T$. Namely, once layers $1,2,...,i-1$ have been learned, layer $i$ will learn the expansion coefficients ($\vec{o}^{(i)}$) and squeeze parameter $r_i$ of the $i$-th most (anti-)squeezed supermode. The network, therefore, discovers the supermodes in their order of significance. For a full circuit ($l=N$), once the entire self-configuring network has converged, the learned circuit transformation satisfies $O_c = O^T$ (up to a multiplication by a diagonal matrix of $\pm 1$ from the left), such that $O_c \Gamma O_c^T = (1-p)K^2 +pI$ reduces to the diagonal form of $\Gamma$ according to Eq. \ref{eq:BMD}. For a sparse network $(l\ll N)$, the circuit routes the $l$ most squeezed supermodes to the outputs, in their order of significance, while the subspace of supermodes orthogonal to the discovered ones remains unprocessed (as depicted in Fig. 1).

Fig. 2 shows a numerical example for learning an entire circuit for the case of $N=10$ modes. In this example, which is valid for both the spatial and spectral implementations described in Fig. 1, the self-configuring layers are learned consecutively. The cost function of Eq. \ref{eq:Rayleigh quotient} is extracted from the homodyne interferogram of Fig. 2a (see Eq. \ref{eq:homodyne} and SI section S2.B) and maximized. In our simulation, we assume a loss probability of $p=0.1$, and model the detection noise due to a final acquisition time (see SI section S2.C). Optimization is performed using automatic differentiation and a variant of stochastic gradient descent \cite{kingma_adam_2017} (see SI, Section~S2)

A formal proof of convergence based on the properties of the Rayleigh quotient is provided in the SI Section S2.D, where it was also shown numerically that each self-configuring layer converges in $O(N)$ iterations. This is evident from the steep convergence of the cost function in learning each of the layers in Fig. 2b. For measuring the performance of the network in recovering the correct BMD, we define a fidelity based on the Hilbert-Schmidt norm of matrices $A$,$B$ of dimension $d$ as $F=\mathrm{Tr} |A^TB|/d$, where $|\cdot|$ corresponds to an element-wise absolute value. As anticipated, the fidelity of the learned orthogonal transformation of the entire circuit $O_C$ with BMD matrix $O^T$ of the ground-truth converges to $1$ after the entire network has been learned, as is also evident by the product $|OO_C|$ depicted in the inset of Fig. 2b.

\begin{figure*}[t]
\centering
\vspace{-0.2cm}
  \includegraphics[trim=0 4cm 0 0, clip, width=\textwidth]{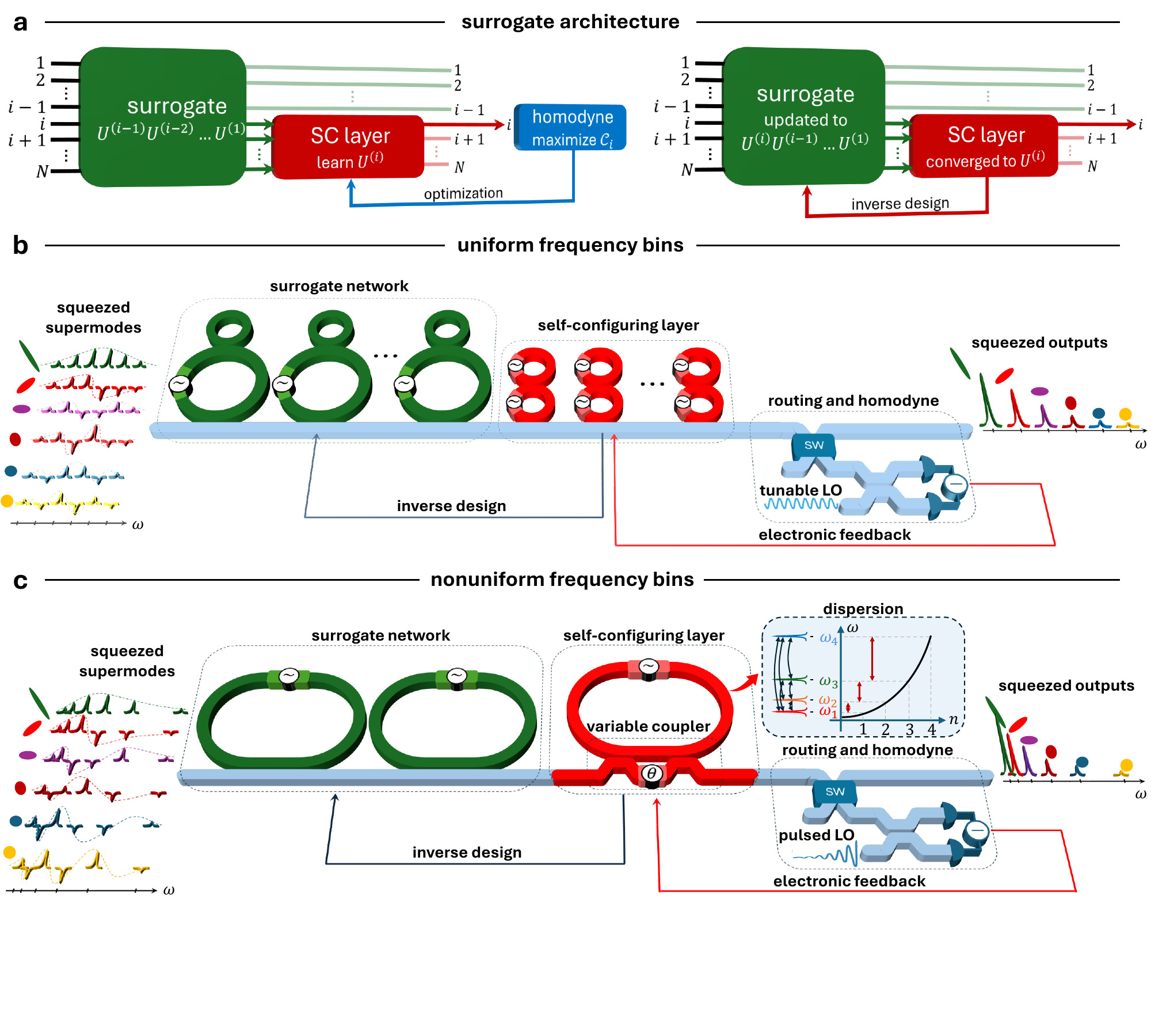}
    \caption{\small \textbf{Frequency domain self-configuring architectures with inverse-designed surrogate networks.} \textbf{a} Block diagram of the circuit, which uses a single self-configuring (SC) layer for learning. After convergence, the surrogate network is inversely-designed to implement the layers learned thus far, as detailed in the main text. \textbf{b} Implementation for uniform frequency bin encoding. The self configuring layer and learning are implemented as in Fig. 1. The surrogate network comprises a cascade of $N$ modulated microring cavities with frequency boundaries (realized by coupling each cavity to a smaller microring). Once a self-configuring layer has been learned, the surrogate netwrok is respectively updated through inverse-design. Overall, the circuit requires $O(N)$ physical elements to implement an entire self-configuring network that discovers all $N$ supermodes. \textbf{c} Implementation for nonuniform frequency bin encoding. Dispersive cavities with non-uniformly spaced resonances are used to process the squeezed supermodes, which are encoded on the same frequency ladder. The self-configuring layer is implemented using intracavity processing in a single modulated cavity with a variable MZI coupler to the waveguide \cite{green_hybrid_2005, yang_low-voltage_2015, yin_high-q-factor_2021}. Once the intracavity processing has concluded, the cavity field is unloaded into the waveguide and measured via homodyne using a pulsed LO, as detailed in the main text and in the SI, Section S3.A. The surrogate network, now comprising only two scattering cavities, is inversely designed accordingly. The circuit now requires $O(1)$ physical elements to implement an entire self-configuring network that discovers all $N$ supermodes. }
    \label{fig:concept}
    \vspace{-0.3cm}
\end{figure*}

\section*{Resource-efficient self-configuring frequency-domain circuits using surrogate networks} 

Frequency-domain quantum photonic circuits - employing synthetic dimensions in photonics \cite{lin_synthetic_2018} - are promising for decreasing the spatial footprint of on-chip circuits. As we show below, it is possible to reduce the number of cavities of a \textit{full network} ($l=N$) to $O(N)$ and even $O(1)$, depending on the implementation. We will begin by describing the main concept and then detail two possible architectures to implement it in the frequency domain. 

The idea is to physically separate the learning of layer $i$ from the implementation of the previously learned layers $1,2,...,i-1$, as depicted in the block-diagram of Fig. 3a. The circuit that performs the learning is a \textit{single} self-configuring layer of maximal length $N-1$ in the frequency domain, as discussed in the previous section. The implementation of the previously learned layers $1,2,...,i-1$ is done using a \textit{surrogate network} situated right before the self-configuring layer that learns layer $i$. The surrogate network is an inverse-designed frequency-domain circuit -- for example, the one proposed in Ref.~\cite{buddhiraju_arbitrary_2021} -- which can be reconfigured after each layer has been learned. 

Explicitly, let us consider the learning of the $i$-th layer of the circuit (represented as a unitary $U^{(i)}$). Before the learning starts,  the surrogate network is first inverse-designed to implement the unitary $U^{(i-1)}U^{(i-2)}...U^{(1)}$ that has been learned thus fur, hence it routes the first $i-1$ supermodes to the first $i-1$ output frequency bins, while the remaining $i,i+1,...,N$ frequency bins carry modes orthogonal to these supermodes. The learning self-configuring layer is then set to have its first $i-1$ MZIs idle (by turning off their modulation and have them act as $2\times 2$ identity), while the optimization is performed over the parameters of the remaining $N-i$ MZIs, trying to maximize the cost function of the $i$-th output frequency bin of Eq.~(\ref{eq:Rayleigh quotient}). Once the learning of the $i$-th layer $U^{(i)}$ has converged, the surrogate network will now be updated, by inverse design, to implement the new unitary $U^{(i)}U^{(i-1)}...U^{(1)}$, and the circuit will be ready to learn the $i+1$-st layer. We now proceed to propose implementations of this concept using two different architectures for frequency-domain quantum processors. 


\textit{Uniform frequency bins.} We start by detailing the first architecture, depicted in Fig. 3b. The self-configuring layer is implemented similarly to the ones in Fig. 1b and contains $N-1$ consecutive frequency-domain MZIs, which can be controllably driven to learn any layer $i=1,...,N$, based on feedback from a homodyne measurement on the frequency bin corresponding to the current layer's output. 

The realization of the surrogate network, implementing the layers $1,...,i-1$, is based on the proposal in Ref. \cite{buddhiraju_arbitrary_2021}. Each of the cavities in the surrogate network is coupled to a smaller ring, forming a finite frequency boundary within which the inverse-designed circuit will operate. Each cavity is also electro-optically driven by a controllable external drive at multiples of the cavity's free spectral range (FSR) $\Omega$. The modulation signal spans multiple tones $\Omega_l = l\Omega$, each with a complex amplitude $\kappa_l\exp(i\phi_l)$, with $\kappa_{-l}=\kappa_l,~\phi_{-l}=-\phi_l$ and $\kappa_0 =0$. The modulation tones $\Omega_l$ thus couple modes $\omega_{n}$ with $\omega_{n\pm l}$ within the bounded frequency range. Assuming a constant photon loss rate (cavity linewidth $\gamma_s$), the unitary scattering matrix of the $j$-th cavity of the surrogate network can be derived from input-output theory as
\begin{equation}
\begin{split}
    \mathrm{U}^{(j)} = \left(\mathrm{H}^{(j)}+i \frac{\gamma_s}{2}\right) \left(\mathrm{H}^{(j)}-i \frac{\gamma_s}{2}\right)^{-1},
\end{split}
    \label{eq:scattering matrix}
\end{equation}
where the coupling matrix $\mathrm{H}^{(j)}_{mn} = \kappa_{m-n}^{(j)} \exp (i\phi_{m-n}^{(j)}) $ is a Hermitian and Toeplitz matrix ($\mathrm{H}^{(j)}_{mn} = \mathrm{H}^{(j)}_{m-n}$) with zero diagonal (or at most constant, e.g., due to electro-optical rectification, self- and cross-phase modulation, in which case it can be calibrated into the cavity Hamiltonian). The coupling matrix thus has $2N-2$ real degrees of freedom. As was numerically proven in Ref. \cite{buddhiraju_arbitrary_2021}, cascading up to $N$ such cavities with the proper inverse design of their coupling matrices allows for the implementation of arbitrary unitary transformations $U=U^{(N)}U^{(N-1)}...U^{(1)}$ with close to unity fidelity. The learning procedure can thus be modeled numerically in an equivalent manner to Fig. 2. Remarkably, this architecture allows for an \textit{entire} circuit to be implemented using exactly $2N$ cavities, instead of requiring $O(N^2)$ in the worst case.

\begin{figure*}[t]
\centering
\vspace{-0.2cm}
  \includegraphics[width=\textwidth]{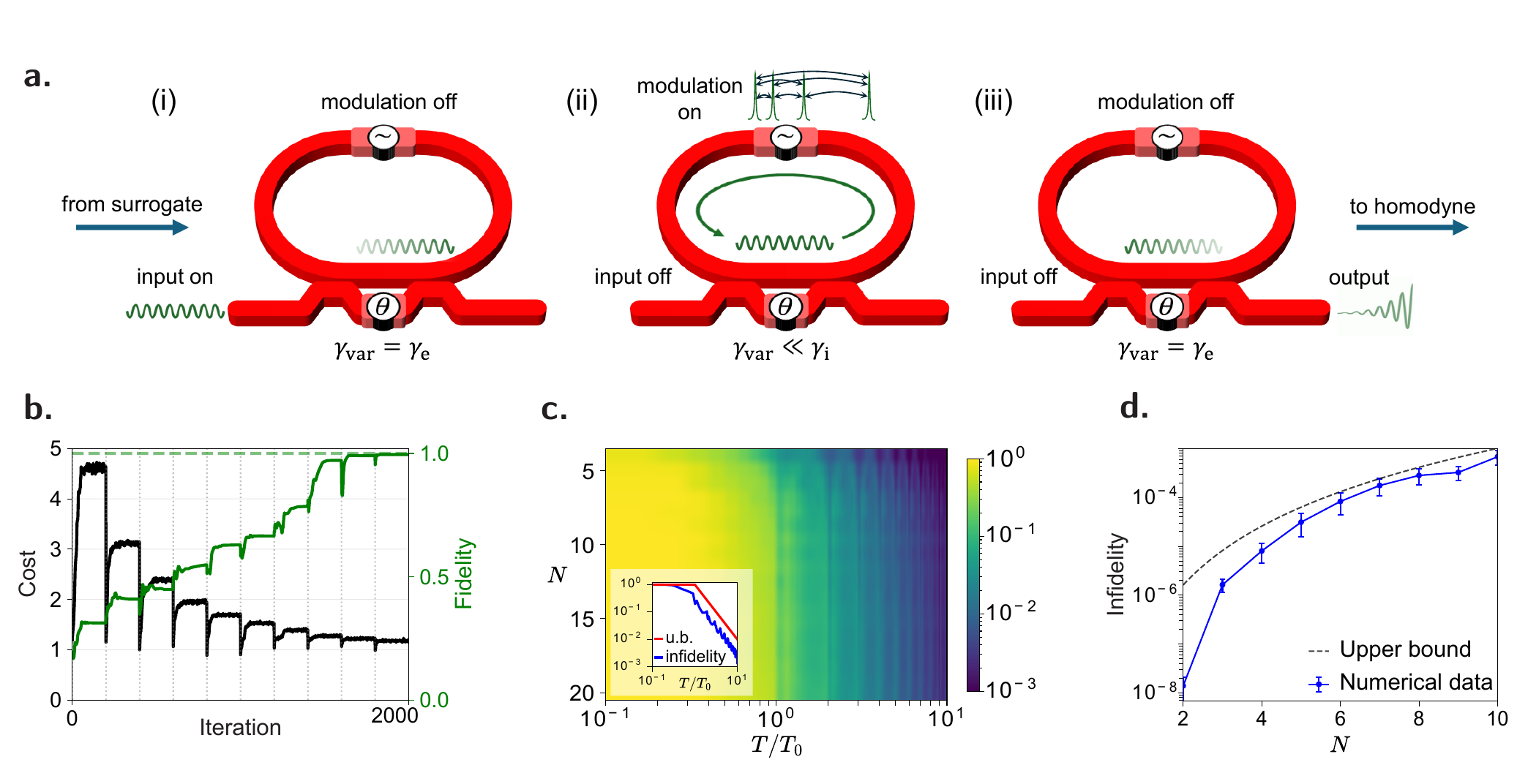}
    \caption{\small \textbf{Variational learning with non-uniform frequency bins.} \textbf{a} Illustration of the intra-cavity learning procedure with the self-configuring cavity, as detailed in the main text. \textbf{b} Simulated learning of the effective covariance $\Gamma_{\mathrm{eff}}$ of Eq. (\ref{eq:Effective covariance matrix}) with $T=10 T_0$, and photon loss probability of $10\%$ for a $N=10$ system. The fidelity is calculated between the circuit $O_{C}$ trained on $\Gamma_{\mathrm{eff}}$ and the BMD orthogonal $O$ of the original input covariance $\Gamma_{\mathrm{in}}(0)$. As the supermodes are weakly morphing, the two converge with a fidelity above $99\%$. For further details, see SI Section S3.B.  \textbf{c} Process infidelity ($1-F$) between the intracavity unitary $U_C = \mathcal{T} \exp \left[-i\int_0^T H(t')dt'\right]$ and the target unitary $U_{C,0} = \exp \left(-iT H_0\right)$, calculated numerically using trotterization, for different values of $N$ and $T/T_0$. The inset shows the upper bound scaling $(T_0/T)^2$. \textbf{d} Process infidelity between the surrogate scattering unitary and the target unitary (calculated using the first-order Magnus expansion; error bars indicate standard deviation over 20 random target unitaries), for the case of quadratic dispersion and a finesse of $\mathcal{F}=1000$, with the upper bound of $(N^2/\pi\mathcal{F})^2$. For more details, see SI Section S3.C-D.  }
    \label{fig:Simulation}
    \vspace{-0.3cm}
\end{figure*}

\textit{Non-uniform frequency bins.} The second architecture we propose employs non-uniformly-spaced frequency bins and dispersive cavities with matching resonances (Fig. 3c). The learning of each self-configuring layer is performed using a single dispersive cavity. In addition, the cavity has a variable coupling to the waveguide (as implemented experimentally in \cite{green_hybrid_2005, yang_low-voltage_2015, yin_high-q-factor_2021}), and the learning is done using intra-cavity processing of the input light over a finite time. The surrogate network, on the other hand, is implemented using two additional scattering cavities. We assume that the multimode squeezed vacuum is a CW signal prepared on the same nonuniform frequency ladder, using, e.g. optical parametric oscillators with a similar dispersion. Before we detail the learning protocol, we provide a few technical aspects on how the frequency-domain mode mixing behaves in this setting.

We consider a quadratic dispersion of the cavities such that the cavity resonances are given by $\omega_n = \omega_0 + n\Omega +n^2 \Omega'$, as depicted in the inset of Fig. 3c. The main conceptual difference is that the cavities are now driven by multiple tones that correspond to \textit{all} possible couplings of frequencies $\omega_n$ and $\omega_m$: $\Omega_{mn} = (m-n)\Omega +(m^2-n^2)\Omega'$, each having a complex amplitude $\kappa_{mn}\exp(i\phi_{mn})$, with $\kappa_{mn}=\kappa_{nm},~\phi_{mn}=-\phi_{nm}$ and $\kappa_{mm}=0$. The resulting coupling matrix now hosts time-dependent detunings. 

The time-dependent coupling matrix can be written as $\mathrm{H}(t)=\mathrm{H}_0 + \delta \mathrm{H}(t)$. Its time-independent part $\mathrm{H}_{0,mn} = \kappa_{mn}\exp(i \phi_{mn})$ corresponds to the desired couplings, forming a dense Hermitian matrix with zero diagonal ($\mathrm{H}_{0,mm} =0$), and supports $N(N-1)$ real degrees of freedom. This stands in contrast to the Toeplitz coupling matrices considered previously for uniform frequency bins. Its time-dependent part $\delta \mathrm{H}(t)$ contains all possible unintentional couplings with their corresponding detunings.

We define the smallest unintentional detuning in $\delta \mathrm{H}(t)$ as $\Delta \equiv \min_{mn\neq jk} |\Omega_{mn} - \Omega_{jk}|$. For a finite modulation time $T$, the unitary transformation generated by $\mathrm{H}(t)$ is $U_C = \mathcal{T} \exp \left[-i \int_0^T \mathrm{H}(t') dt'\right]$ with $\mathcal{T}$ denoting time ordering. In the limit where $T \gg 2\pi/\Delta \equiv T_0$, the time dependent term $\delta \mathrm{H} (t)$ is fast oscillating and can be dropped, and $U_C$ approaches the target unitary $U_{C,0} \equiv \exp \left(-i \mathrm{H}_0T \right)$ (we further discuss the validity of this approximation below, as well as in the SI Sections S3.C-D).

We consider the learning of the first supermode, such that the surrogate network (depicted in green in Fig. 3c) is turned off (not driven). The learning protocol of the self-configuring cavity (depicted in red in Fig. 3c) proceeds as follows, and depicted in Fig. 4a. First, the self-configuring cavity is initially empty and not driven. The coupler is then varied to a maximal external coupling to the waveguide, $\gamma_e$, such that the cavity is coupled to the CW squeezed vacuum input and steady-state is reached. Then, the coupler is varied again to minimize the coupling with the waveguide (to a value smaller than the intrinsic loss rate $\gamma_i$). The CW input is turned off. Second, the cavity is then modulated for a finite modulation time $T\gg T_0$ with a target coupling matrix $\mathrm{H}_0$, to implement a self-configuring layer $O_C$ (corresponding to the target unitary $U_{C,0}$) that acts on the quantum light trapped in the cavity. Third, the coupler is reset to $\gamma_e$ to dump the processed light into the waveguide. The resulting output signal takes a pulsed waveform. For this reason, to perform a homodyne measurement on the first frequency bin $\omega_0$, the output signal is combined with a \textit{pulsed} LO with the same temporal shape, $e^{-i\omega_0 t}\alpha_{LO}(t)$, unlike the frequency-domain architectures of Figs. 1b and 3b, which employed a CW LO. Each such measurement is performed for a single LO phase $\phi$, and these steps are repeated until a full interferogram is acquired. For more details on this four-step learning procedure, see SI Section S3.A.

In the SI, Section S3.B, we show that for a maximal temporal overlap between the pulsed LO and the pulsed output signal, this entire protocol diagonalizes an effective covariance matrix given by
\begin{equation}
\begin{split}
\Gamma_{\mathrm{eff}} &= e^{-\gamma_i T}  \int d\omega \frac{1}{\pi} \frac{\gamma_e/2}{(\gamma_e/2)^2+\omega^2}\Gamma_{\mathrm{in}}(\omega) \\&+ (1-e^{-\gamma_i T})\mathrm{I},
\end{split}
    \label{eq:Effective covariance matrix}
\end{equation}
where $\Gamma_{\mathrm{in}}(\omega)$ is the frequency-dependent covariance matrix of the original CW input, and where $1-e^{-\gamma_i T}$ serves as the photon loss probability, producing a similar form to Eq. (\ref{eq:BMD}). In the SI Section S3.B, we show that under our assumptions, $\Gamma_{\mathrm{eff}}$ shares the same supermodes and eigenvalue ordering as the input covariance matrix. 

The surrogate network (depicted in green in Fig. 3c) comprises two scattering cavities with the same non-uniform dispersion, and we denote their linewidth as $\gamma_s$. Similarly to the self-configuring cavity, the scattering cavities are driven with time-dependent coupling matrices $\mathrm{H}^{(j)}(t)$ (with $j=1,2$), which, in the limit of $\gamma_s^{-1}\gg T_0$ approach the time-independent target matrices $\mathrm{H}^{(j)}_0$, and the surrogate network implements the target unitary $U_{C,0}=U_{C,0}^{(2)}U_{C,0}^{(1)}$, where the $U_{C,0}^{(j)}$ are given by Eq. (\ref{eq:scattering matrix}) with $H^{(j)}=H^{(j)}_0$. Generalizing Ref. \cite{buddhiraju_arbitrary_2021}, in the SI Section S3.C, we show that two such cavities are sufficient for the inverse design of an arbitrary unitary. Importantly, once the learning has been concluded, the surrogate network can scatter the incident CW input to a CW output, decomposed into the squeezed supermodes. 

Fig. 4b depicts a numerical simulation of the learning with non-uniform frequency bins, for $T=10T_0$, $N=10$ modes and $10\%$ photon loss probability. For $\Gamma_{\mathrm{in}}(\omega)$, we emulate a physical input prepared using an OPO squeezer followed by a mode-mixing cavity (for details, see SI Section S3.B), and let the network learn $\Gamma_{\mathrm{eff}}$ of Eq. (\ref{eq:Effective covariance matrix}). In the considered regime, the supermodes are weakly morphing, and the circuit trained on $\Gamma_{\mathrm{eff}}$ learns the BMD of $\Gamma_{\mathrm{in}}$ with a fidelity of $99.58\%$, as shown in Fig. 4b.

We note that imperfect fidelity between the target and actual unitaries $U_{C}$ and $U_{C,0}$, of both the self-configuring and surrogate cavities, can affect the performance of the learning. We consider the process fidelity \cite{nielsen_simple_2002} between these two unitaries $F=|\mathrm{Tr}[U^{\dagger}_C U_{C,0}]|^2/N^2$ as a function of the relative modulation time $T/T_0$, the cavity's lifetime, and the mode number $N$, which sets a limit to the mode capacity of this architecture. In the SI Section S3.D, we show that in general the process infidelity scales as $1-F \propto (T_0/T)^2$. Specifically, for the case of quadratic dispersion, we have $1-F \propto (N^2/\mathcal{F})^2$, where $\mathcal{F}$ denotes the cavity finesse. We confirm this scaling numerically for both the intracavity processing unitary as well as the surrogate scattering matrix, as depicted in Figs. 4c and 4d, respectively.    

\section*{Discussion}
We proposed and analyzed scalable on-chip architectures for processing multi-mode squeezed light. Using the notion of self-configuring networks, the photonic circuit can be sequentially optimized using homodyne measurements, which define an observable cost function that is then maximized. This method is a manifestation of the variational principle, and enables us to learn and route the most squeezed supermodes of an input quantum state in their order of significance. 

We discussed implementations in both the spatial and spectral domains using real and synthetic meshes of Mach-Zehnder interferometers (MZIs), allowing one to reduce the spatial footprint of such quantum processing units. Sparse networks - which discover the first $l\ll N$ dominant supermodes, where $N$ is the number of modes, can be implemented using $O(lN)$ MZIs instead of $O(N^2)$ for a full network. In the frequency domain, the spatial footprint can be further reduced by employing inverse-designed surrogate networks that emulate the circuit learned thus far. Using two different frequency bin encoding schemes, we showed that the number of physical elements needed to implement an entire network (learning all $N$ supermodes) can be reduced to $O(N)$ and even $O(1)$.

Our methods can inspire further development in multimode continuous-variable quantum technologies. An immediate application of our methods could be quantum-enhanced sensing across distributed channels. Once the supermodes are efficiently demultiplexed, the most squeezed outputs can be used as probes for different systems, as part of the distributed quantum sensing protocol \cite{zhuang_distributed_2018,guo_distributed_2020}. Another avenue for quantum metrology is the use of the self-configuring methods for extracting the quantum Fisher information of the generated quantum light \cite{abbasgholinejad_theory_2025}.

Our results could be especially exciting for the emerging field of frequency-bin quantum information processing \cite{lu_frequency-bin_2023}, where architectures for on-chip quantum processing units are currently being explored. We envision that self-configuring networks could also be useful for characterizing frequency-domain graph states, allowing scalable measurement of the entanglement witnesses (nullifiers) of these high-dimensional states \cite{pfister_continuous-variable_2019}.

The ideas presented in this work could be extended to other encoding schemes. For example, it will be interesting to extend our protocols to time-bin and hybrid frequency-time bin encodings \cite{miller_analyzing_2020}, which have been shown useful for scaling up high-dimensional entanglement \cite{erhard_advances_2020,kues_quantum_2019}. Moreover, it can be useful to explore other non-uniform frequency bin encodings (such as the Golomb ruler \cite{golomb_how_1972}) to further improve circuit fidelity and mode capacity. Finally, our methods could generalize to other forms of multimode quantum noise measurement, such as in the photon-number basis and higher-order intensity correlations, for studying unique multimode and nonlinear quantum optical systems \cite{rivera_ultra-broadband_2025,zia_uddin_noise-immune_2025}.


\section{Funding}
This project was funded by Toyota Research Institute of North America, grant XXX. A.K. is supported by the VATAT-Quantum fellowship by the Israel Council for Higher Education; the Urbanek-Chodorow postdoctoral fellowship by the Department of Applied Physics at Stanford University; the Zuckerman STEM leadership postdoctoral program; and the Viterbi fellowship by the Technion. C.~R.-C. is supported by a Stanford Science Fellowship. S.~F. and D.A.B.M. acknowledge support by the Air Force Office of Scientific Research (AFOSR, grant FA9550-21-1-0312). D.A.B.M. also acknowledges support by the Air Force Office of Scientific Research (AFOSR, grant FA9550-23-1-0307). E.L., J.S. and J.V. were supported by the DARPA QUICC program and AFOSR award no. FA9550-23-1-0248.

\section{Authors' contributions}

A.K., C.R.C. and P.-A.M. contributed equally to this work.

\section{Competing interests statement}
The authors are seeking patent protection for invention related to this work (US Patent Application XXX, Stanford University). 

\section{Acknowledgements}

The authors acknowledge Kejie Fang and Toyota Research Institute of North America for useful discussions.

\bibliographystyle{ieeetr}
\bibliography{references}

\end{document}